\documentclass[12pt]{article}

\usepackage{amsmath,amssymb,a4,epsfig,feynmf}

\allowdisplaybreaks[4]
\renewcommand{\title}[1]{\null\vspace{10mm}\noindent
                         {\Large{\bf #1}}\vspace{10mm}}
\newcommand{\authors}[1]{\noindent{\large #1}\vspace{20mm}}
\newcommand{\address}[1]{{\center{\noindent\small\itshape #1\vspace{0mm}}}}

\makeatletter
\def\section{\@startsection{section}{1}{\z@}{-3.25ex plus -1ex minus
             -.2ex}{1.5ex plus .2ex}{\normalfont\bfseries}}
\def\subsection{\@startsection{subsection}{1}{\z@}{-3.25ex plus -1ex
                minus -.2ex}{1.5ex plus .2ex}{\normalfont\itshape}}

\renewenvironment{thebibliography}[1]
         {\section*{References}\frenchspacing\small
          \begin{list}{[\arabic{enumi}]}
         {\usecounter{enumi}\parsep=2pt\topsep 0pt
         \settowidth{\labelwidth}{[#1]}
         \leftmargin=\labelwidth\advance\leftmargin\labelsep
         \rightmargin=0pt\itemsep=0pt\sloppy}}{\end{list}}
\makeatother

\begin{document}

\begin{titlepage}
\thispagestyle{empty}

\begin{center}
\hspace*{\fill}{{\normalsize hep-th/0206018-v3}}

\title{
Quantum field theories on noncommutative $\mathbb{R}^4$ 
\\[2mm]
versus $\theta$-expanded quantum field theories}

\authors {Raimar Wulkenhaar$\,^*$}    \vspace{-20mm}

\address{Max-Planck-Institut f\"ur Mathematik in den 
Naturwissenschaften\\
Inselstra\ss{}e 22--26, D-04103 Leipzig, Germany}

{\renewcommand{\thefootnote}{\fnsymbol{footnote}}
\footnotetext[1]{raimar.wulkenhaar@mis.mpg.de, Schloe\ss{}mann fellow}
}   

\vskip 3cm

\begin{minipage}{12cm}
  
  {\it Abstract.} I recall the main motivation to study quantum field
  theories on noncommutative spaces and comment on the most-studied
  example, the noncommutative $\mathbb{R}^4$. That algebra is given by
  the $\star$-product which can be written in (at least) two ways: in
  an integral form or an exponential form. These two forms of the
  $\star$-product are adapted to different classes of functions,
  which, when using them to formulate field theory, lead to two
  versions of quantum field theories on noncommutative $\mathbb{R}^4$.
  The integral form requires functions of rapid decay and a
  (preferably smooth) cut-off in the path integral, which therefore
  should be evaluated by exact renormalisation group methods.  The
  exponential form is adapted to analytic functions with arbitrary
  behaviour at infinity, so that Feynman graphs can be used to compute
  the path integral (without cut-off) perturbatively. 

\end{minipage}

\end{center}
\end{titlepage}

\setcounter{section}{-1}

\section{Disclaimer}

This is not a review. The organisers of the Hesselberg 2002 workshop
on ``Theory of renormalisation and regularisation'', Ryszard Nest,
Florian Scheck and Elmar Vogt, asked me to present something about
$\theta$-deformed quantum field theories and to prepare some notes for
the proceedings. In the following I will present the logic behind and
the results of my own work on this subject. Objectivity and
completeness are not the aim of this presentation. I have quoted
references where I knew of them, statements without citation do not
mean that they are new.

These notes go far beyond my presentation at the Hesselberg workshop.
They reflect my current point of view\footnote{Changes in 
  v2 are due to an e-mail exchange with Mohammad Sheikh-Jabbari on the
  different $\star$-products and discussions with Dorothea Bahns and
  Klaus Fredenhagen who, in particular, convinced me that Minkowskian
  noncommutative field theories are different.\\ \hspace*{\parindent}
  Changes in v3 go back to very useful comments by Edwin Langmann who
  explained to me that the two versions of the $\star$-product which
  in the previous versions were regarded as different products are
  actually two extensions of \emph{the same product} to different
  classes of functions.}, the formulation of which evolved through the
typing of papers to be found in the hep-th arXiv and the preparation
of invited talks for workshops, conferences and invitations in Wien,
Nottingham, Jena, Marseille, Leipzig, M\"unchen, Hesselberg, Leipzig
(again), Oberwolfach, Hamburg and Trieste (Wien will follow). I am
grateful to the organisers of these events for invitation and
hospitality, as well as to my friends for discussions and cooperation.

\section{Farewell to manifolds}
\label{farewell}

Half a century of high energy physics has drawn the following picture
of the microscopic world: There are matter fields and carriers
of interactions between them. Four different types of interactions
exist: electromagnetic, weak and strong interactions as well as
gravity. The traditional mathematical language to describe these
structures of physics is that of fibre bundles. The base manifold $M$
of these bundles is a four-dimensional metric space with line element 
$ds^2= g_{\mu\nu}(x)\, dx^\mu dx^\nu$. Matter fields $\psi$
are sections of a vector bundle $V$ over $M$. The carriers of
electromagnetic, weak and strong interactions are described by
connection one-forms $A$ of $U(1)$, $SU(2)$ and $SU(3)$ principal fibre
bundles, respectively. Gravity is the determination of the metric
$g$ by the one-forms $A$ and 
sections $\psi$, and vice-versa.

The dynamics of $(A,\psi,g)$ is governed by an action
functional $\Gamma[A,\psi,g]$, which yields the equations of motions
when varied with respect to $A,\psi,g$. The complete action functional
for the phenomenologically most successful model, the standard model
of particle physics, is an ugly patchwork of unrelated pieces when
expressed in terms of $(A,\psi,g)$. 

Next there is a clever calculus, called \emph{quantum field theory},
which as the input takes the action functional $\Gamma$ and as the
output returns \emph{numbers}. On the other hand, there are
experiments which also produce numbers. There is a remarkable
agreement\footnote{There are of course experimental data which so far
  could not be reproduced theoretically, such as the energy spectrum of
  hadrons.}  of up to $10^{-11}$ between corresponding numbers
calculated by quantum field theory and those coming from experiment.
This tells us two things: The action functional (here: of the standard
model) is very well chosen and, in particular, quantum field theory is
an extraordinarily successful calculus.

There is however, apart from the description of strong interactions at
low energy, a tiny problem: one of the basic assumptions of quantum
field theory is not realised in nature. First, the metric $g$ is
considered in quantum field theory as an external parameter,
and---mostly---the calculus works only if the metric is that of
Euclidian or Minkowski space, $g_{\mu\nu}=\delta_{\mu\nu}$ or
$g_{\mu\nu}=\mathrm{diag}(1,-1,-1,-1)$, respectively. But let us
ignore this and assume for a moment that quantum field theory works on
any (pseudo-) Riemannian manifold. Let us then ask how we measure
technically the geometry.  The building blocks of a manifold are the
\emph{points} labelled by coordinates $\{x^\mu\}$ in a given chart.
Points enter quantum field theory via the sections $\psi(x)$ and
$A(x)$, i.e.\ the \emph{values} of the fields at the point labelled by
$\{x^\mu\}$. This observation provides a way to ``visualize'' the
points: we have to prepare a distribution of matter which is sharply
localised around $\{x^\mu\}$. For a perfect visualisation we need a
$\delta$-distribution of the matter field. This is physically not
possible, but one would think that a $\delta$-distribution could be
arbitrarily well approximated. However, that is not the case, there
are limits of localisability long before the $\delta$-distribution is
reached \cite{Doplicher:tu}.

Let us assume there is a matter distribution which is believed to have
two separated peaks within a space-time region $R$ of diameter
$\mathrm{d}$.  How do we test this conjecture? We perform a scattering
experiment in the hope to find interferences which tell us about the
internal structure in the region $R$. We clearly need test particles
of de Broglie wave length $\lambda =\frac{\hbar c}{E}\lesssim
\mathrm{d}$, otherwise we observe a single peak even if there is a
double peak. For $\lambda \to 0$ the gravitational field of the test
particles becomes important. The gravitational field created by an
energy $E$ can be measured in terms of the Schwarzschild radius
\begin{align}
r_s = \frac{2 G E}{c^4} = \frac{2 G \hbar}{\lambda c^3} \gtrsim 
\frac{2 G \hbar}{\mathrm{d} c^3}\; ,   
\end{align}
where $G$ is Newton's constant. If the Schwarzschild radius $r_s$
becomes larger than the radius $\frac{\mathrm{d}}{2}$, the inner
structure of the region $R$ can no longer be resolved (it is behind
the horizon). Thus, $\frac{\mathrm{d}}{2} \geq r_s$ leads to the
condition
\begin{align}
\frac{\mathrm{d}}{2} \gtrsim \ell_P := \sqrt{\frac{G \hbar}{c^3}}\;,
\end{align}
which means that the Planck length $\ell_P$ is the fundamental length
scale below of which length measurements become meaningless.
Space-time cannot be a manifold.

\section{Spectral triples}

What does this mean for quantum field theory? It means that we cannot
trust traditional quantum field theories like the (quantum) standard
model because they rely on \emph{non-existing information} about the
short-distance structure of physics which determines the loop
calculations.

What else can we take for space-time? A lattice? The disadvantage of
the lattice is that \emph{symmetries}, which are guiding principles in
quantum field theory, are lost. There are also problems with the spin
structure. Lattice calculations are regarded as a
mathematically rigorous method, but at the end mostly the continuum
limit is desired in which the symmetries are intended to be restored.

The lattice approach points into the right direction. A lattice is a
metric space but not a differentiable manifold. What we would like to
have as candidates for space-time is a class of metric spaces which
are equipped with a differential calculus and, additionally, a spin
structure to allow for fermions. Such objects exist in mathematics,
they are called \emph{spectral triples} \cite{Connes:tu,
  Connes:1996gi}. They are \emph{noncommutative geometries}
\cite{Connes, Gracia-Bondia:tr} which are the closest generalisation
of differentiable spin manifolds. There are good reasons to believe
that spectral triples are the right framework for physics.
\begin{enumerate}
\item The language in terms of which spectral
triples are formulated comes from field theory: Besides the algebra
$\mathcal{A}$ represented on a Hilbert space $\mathcal{H}$ (which
alone are only good for measure theory), to describe \emph{metric
  spaces with spin structure} one also needs a Dirac operator
$\mathcal{D}$, the chirality $\gamma^5$ and the charge conjugation
$J$, see \cite{Connes:1996gi}. 

\item The standard model of particle
physics looks much simpler when formulated in the language of spectral
triples\footnote{In fact the axioms of spectral triples
  \cite{Connes:1996gi} are tailored such that the (Euclidian) standard
  model is a spectral triple.}.  This is first of all due to the
understanding of the Higgs field as a component of a gauge field
living on a spectral triple. The $(\phi^4- m^2 \phi^2)$ Higgs
potential comes from the same source as the Maxwell Lagrangian
$F_{\mu\nu} F^{\mu\nu}$, and the Yukawa coupling of the Higgs with
the fermions has the same origin as the minimal coupling of the gauge
fields with the fermions. But the connection is much deeper, for
instance, the spectral triple description \emph{enforces} the
following (in the language of Yang-Mills-Higgs models unrelated)
features \cite{Carminati:1997ej}:
\\
\begin{tabular}{rl}
(i) & weak interactions break parity maximally\\
(ii) & weak interactions suffer spontaneous breakdown\\
(iii) & strong interactions do not break parity \\
(iv) & strong interactions do not suffer spontaneous breakdown
\end{tabular}

\item The separation of gauge fields and gravity starts to
disappear: Yang-Mills fields, Higgs fields and gravitons are all
regarded as \emph{fluctuations} of the free Dirac operator
\cite{Connes:1996gi}. The spectral action
\begin{align}
\Gamma &= \mathrm{trace}
\,\chi \Big(z\frac{\mathcal{D}^2}{\Lambda^2}\Big)\;,
& 
\chi(t) = \left\{ \begin{array}{cl} 1 & \text{ for } 0\leq t \leq 1 
\\ 
0 & \text{ for } t>1 
\end{array} \right.
\label{spectral}
\end{align}
(which is the weighted sum of the eigenvalues of $\mathcal{D}^2$ up to
the cut-off $\Lambda^2$) of the single fluctuated Dirac operator
$\mathcal{D}$ gives the complete bosonic action of the standard model,
the Einstein-Hilbert action (with cosmological constant) and an
additional Weyl action term in one stroke \cite{Chamseddine:1996zu}.
The parameter $z$ in (\ref{spectral}) is the ``noncommutative coupling
constant'' \cite{Carminati:1997ej}. Assuming the spectral action
(\ref{spectral}) to produce the bare action at the (grand unification)
energy scale $\Lambda$, the renormalisation group equation based on
the one-loop $\beta$-functions leads to a Higgs mass of $182\dots
201\,\mathrm{GeV}$ \cite{Carminati:1997ej}.

\end{enumerate}

There are of course technical difficulties with spectral triples, such
as the restriction to compact spaces with Euclidian signature, but it
is clear that spectral triples are a very promising strategy. For
attempts to overcome Euclidian signature see \cite{Kopf, Strohmaier}
and for an extension to non-compact spaces \cite{Gracia-Bondia:2001ct}.

The strength of the spectral triple approach is that it leads
immediately to classical action functionals with a lot of symmetries,
even on spaces other than manifolds. We can feed the spectral action
functional into our calculus of quantum field theory in order to
produce numbers to be compared with experiments. One of the
formulations of that calculus, the path integral approach, is
perfectly adapted to spectral triples. All one needs are \emph{labels}
$\Phi$ for the degrees of freedom of the spectral action
$\Gamma_{cl}[\Phi]$ in order to write down (at least formally) the
measure $\mathcal{D}[\Phi]$ for the (Euclidian) path integral
\begin{align}
Z[J] = \int \mathcal{D}[\Phi] \,\exp\Big(-\frac{1}{\hbar}
\Gamma_{cl}[\Phi] - \big\langle J,\Phi \big\rangle\Big)\;.
\label{pathintegral}
\end{align}
The source $J$ is an appropriate element of the dual space of the
$\Phi$'s.  Everything interesting (in a Euclidian quantum field
theory) can be computed out of $Z[J]$. It is not important how one
labels the degrees of freedom, because $Z[J]$ is invariant under a
change of variables \cite{Kamefuchi:sb}. 

However, we cannot completely exclude the possibility that quantum
field theory is implicitly built upon the assumption that the action
functional taken as input lives on a manifold.  The best way to test
whether the standard calculus of quantum field theory extends to
spectral triples is to apply it to examples which are
\emph{deformations} of a manifold. Let us assume there is a family of
spectral triples which are distinguished by a set of parameters
$\theta$ such that for $\theta \to 0$ we recover an ordinary manifold.
Then we should expect that the family of numbers computed out of
(\ref{pathintegral}) for any $\theta$ tends for $\theta \to 0$ to the
numbers computed for the manifold case.  Otherwise something is wrong.
It is unlikely that the problem (if any) lies in the formula
(\ref{pathintegral}) itself, which is very appealing. However, the
evaluation of (\ref{pathintegral}) often involves formal manipulations
which may work in one case but fail in another one. We should be
careful.

I should mention that the situation is much more difficult in the case
of Minkowskian signature of the metric. Apart from the difficulty to
extend the definition of spectral triples to geometries with
non-Euclidian signature and the mathematical problems of the
non-Euclidian path integral, there is evidence now \cite{Bahns:2002vm}
that formal Wick rotation in the Feynman graph computations based on
the path integral (\ref{pathintegral}) does not yield the correct
theory in the noncommutative case.

Before going to the example let us remind ourselves what the challenge
was. We need a replacement for the space-time manifold which is not
based on the notion of points. The replacement is expected to be a
spectral triple, but in order to compare the outcome with experiments,
we have to be sure that the calculus of quantum field theory can be
applied. This is why we are interested in spectral geometries other
than manifolds on which quantum field theoretical computations are
\emph{possible to perform}. We do not claim that our examples are
the correct description of the real world.

\section{The noncommutative torus}

It is time for an example. The simplest noncommutative spectral triple
is the noncommutative $d$-torus, see e.g.\ \cite{Rieffel}. A basis for
the algebra $\mathbb{T}^d_\theta$ of the noncommutative $d$-torus is
given by unitarities $U^p$ labelled by $p=\{p_\mu\}\in \mathbb{Z}^d$,
with $U^p (U^p)^* = (U^p)^* U^p =1$. The multiplication is defined by
\begin{align}
U^p U^q = \mathrm{e}^{\mathrm{i} \pi \theta^{\mu\nu} p_\mu q_\nu} 
U^{p+q} \;,\qquad \mu,\nu=1,\dots,d\;,\quad \theta^{\mu\nu}
=-\theta^{\nu\mu} \in \mathbb{R}\;. 
\end{align}
Elements $a\in \mathbb{T}^d_\theta$ have the following form:
\begin{align}
a = \sum_{p \in \mathbb{Z}^d} a_p U^p\;,\qquad a_p \in
\mathbb{C}\;,\quad \|p\|^n |a_p| \to 0 \text{ for } \|p\| \to \infty\;.
\label{Atheta}
\end{align}
If $\theta^{\mu\nu} \notin \mathbb{Q}$ (rational numbers) one can
define partial derivatives 
\begin{align}
\partial_\mu U^p := -\mathrm{i} p_\mu U^p\;,
\label{pd}
\end{align}
which satisfy the Leibniz rule and Stokes' law with respect to the
integral 
\begin{align}
\int a = a_0\;,  
\label{trace}
\end{align}
where $a$ is given by (\ref{Atheta}). The algebra $\mathbb{T}^d_\theta$
gives rise to a Hilbert space by GNS construction with respect to
(\ref{trace}), and the partial derivatives (\ref{pd}) yield a Dirac
operator. Algebra, Hilbert space and Dirac operator extend to a
spectral triple satisfying all axioms. For details (and a discussion
of the rational case $\theta^{\mu\nu} \in \mathbb{Q}$) see
\cite{Krajewski:1998gq}. The noncommutative torus was the first
noncommutative space where field theory has been studied
\cite{ConnesRieffel}. 

The spectral action (\ref{spectral}) for this spectral triple reads 
\begin{align}
\Gamma = \frac{1}{4 g^2} \int F_{\mu\nu} F^{\mu\nu}\;,\qquad 
F_{\mu\nu} = \partial_\mu A_\nu -  \partial_\nu A_\mu - \mathrm{i}
(A_\mu A_\nu-A_\nu A_\mu)\;,
\label{actiontorus}
\end{align}
where $A_\mu=A_\mu^* \in \mathbb{T}^d_\theta$. Then the path integral
(\ref{pathintegral}) is evaluated in terms of Feynman graphs, which
involve \emph{sums}, not integrals, over the discrete loop momenta.
At the one-loop level it is possible to extract the pole parts of
these sums via $\zeta$-function techniques \cite{Krajewski:1999ja}. The
result is that the quantum field theory associated to the classical
action (\ref{actiontorus}) is divergent but (for $d=4$ dimensions)
one-loop renormalisable (divergences are multiplicatively removable
and the Ward identities are satisfied) \cite{Krajewski:1999ja}. 

Everything is perfect so far. Unfortunately, nobody was able to
investigate this model at two and more loops, for the simple reason
that sums are more difficult to evaluate than integrals. It is
dangerous here to approximate the sums by integrals, because the critical
question is the behaviour at small $p$, see below. The most important
property of the torus is that the zero mode $p=0$ decouples, and the
next-to-zero modes are ``far away'' from zero $(p \in \mathbb{Z}^d)$.
Taking $p\in \mathbb{R}^d$ to deal with integrals, the zero mode
decouples as well, but the next-to-zero modes are infinitesimally
close to zero. Due to the ease of the computations, much more work has
been performed on the non-compact analogue of the noncommutative
$4$-torus---the noncommutative $\mathbb{R}^4$. 

\section{The noncommutative $\mathbb{R}^4$}

Therefore, let us pass to the noncommutative $\mathbb{R}^4$. The
algebra $\mathbb{R}^4_\theta$ is given by the space
$\mathcal{S}(\mathbb{R}^4)$ of (piecewise) Schwartz class functions of
rapid decay, equipped with the multiplication rule
\cite{Gracia-Bondia:1987kw}
\begin{align}
(a\star b)(x) &= \int \frac{d^4k}{(2\pi)^4} \int d^4 y \; 
a(x{+}\tfrac{1}{2} \theta {\cdot} k)\, b(x{+}y)\,
\mathrm{e}^{\mathrm{i} k \cdot y}\;,
\label{starprod}
\\*
&
 (\theta {\cdot} k)^\mu = \theta^{\mu\nu} k_\nu\;,\quad
k{\cdot}y = k_\mu y^\mu\;,\quad \theta^{\mu\nu}=-\theta^{\nu\mu}\;.
\nonumber
\end{align}
There is again a $\theta$, which however is completely different from
the one in (\ref{Atheta}). The entries $\theta^{\mu\nu}$ in
(\ref{starprod}) have the \emph{dimension} of an area whereas in
(\ref{Atheta}) they are numbers (for the torus everything can be
measured in terms of the radii).  There is also no rational situation
for the $\mathbb{R}^4_\theta$. Note that the product (\ref{starprod}) is
associative but noncommutative, and $(a\star
b)(x)\big|_{\theta=0}=a(x) b(x)$. 

It is interesting to perform a Taylor expansion of (\ref{starprod}) about
$\theta=0$: 
\begin{align}
(a\star_\omega b)(x) &:= \sum_{n=0}^\infty \frac{1}{n!} 
\theta^{\alpha_1 \beta_1} \cdots \theta^{\alpha_n \beta_n} 
\Big(\frac{\partial^n
(a\star b)(x)}{\partial \theta^{\alpha_1 \beta_1} \dots 
\partial \theta^{\alpha_n \beta_n}}\Big)_{\theta=0}
\nonumber
\\*
&= \Big(\mathrm{e}^{\frac{\mathrm{i}}{2} \theta^{\alpha\beta}
  \frac{\partial}{\partial y^\alpha}
  \frac{\partial}{\partial z^\beta}} \;a(x{+}y)\, b(x{+}z)
\Big)_{z=y=0}\;.
\label{staromega}
\end{align}
What is the relation between $\star$ and $\star_\omega$? I first
thought that they are completely different products. I am grateful to
Edwin Langmann for explaining to me that $\star$ and $\star_\omega$
are actually the same products, the point is that the derivatives in
(\ref{staromega}) are actually generalised derivatives in the sense of
distribution theory. There is a class of functions on which $\star$
and $\star_\omega$ (with the derivatives taken literally) coincide,
these are analytic functions of rapid decay. Then, depending on how
one extends the class of functions to less regular ones, different
forms for displaying the product are preferred. 

The $\star$-product (\ref{starprod}) has excellent smoothing
properties, and the multiplier algebra $\mathcal{M}_\theta$, the set
of all distributions which when $\star$-multiplied with elements of
$\mathbb{R}^4_\theta$ give again elements of $\mathbb{R}^4_\theta$, is
very big \cite{Gracia-Bondia:1987kw}. The $\star$-product is clearly
\emph{non-local}: to the value of $a\star b$ at $x$ there contribute
values of $a,b$ at points far away from $x$. The form (\ref{starprod})
is very convenient for piecewise Schwartz class functions, in
particular for functions of compact support, as the computation of a
two-dimensional example in Appendix A shows. This calculation shows
that the $\star$-product has some surprising (at least to me)
behaviour at very short distances. Very similar calculations have
already been performed in \cite{Bars:2001dy}, with similar
conclusions. The most impressive behaviour is shown in Figure 3. The
$\star$-product completely smoothes away the (``extremely-localised''
\cite{Bars:2001dy}) modes of support within an area $\theta$.  It is
apparent that $\theta$ acts as a cut-off (or a horizon in terms of
gravitational physics), a cut-off which preserves all symmetries!
Thus, the $\mathbb{R}^4_\theta$ with $\star$-product (\ref{starprod})
is an excellent model for space-time.

The focus of the $\star_\omega$-product is a different one. In order
to interpret the partial derivatives literally, one has to stay
within the class of analytic functions. However, there is no need of
rapid decay at infinity. The $\star_\omega$-product is e.g.\ defined
for polynomials of finite degree or for plane waves:
\begin{align}
\mathrm{e}^{\mathrm{i}p_\mu x^\mu} \star_\omega  
\mathrm{e}^{\mathrm{i}q_\nu x^\nu} = 
\mathrm{e}^{-\frac{\mathrm{i}}{2}\theta^{\alpha\beta} 
p_\alpha q_\beta }\,\mathrm{e}^{\mathrm{i}(p_\mu+q_\mu) x^\mu}\;.
\end{align}
There is no need to assemble the plane waves to wave packets of rapid
decay at infinity. The non-locality of the $\star_\omega$-product
(\ref{staromega}) is hidden. To the value of $a\star_\omega
b$ at $x$ there contribute values of $a,b$ in an infinitesimal
neighbourhood of $x$ only, but taking the derivatives literally
requires $a,b$ to be analytic, which means that the infinitesimal
neighbourhood of $x$ contains all information about $a,b$ on the
entire $\mathbb{R}^4$.


\section{The geometry of $\mathbb{R}^4_\theta$}
\label{spec}

In my opinion a thorough investigation of the spectral geometry of
$(\mathbb{R}^4_\theta,\mathcal{H},\mathcal{D})$ must precede any
quantum field theoretical computations of models on
$\mathbb{R}^4_\theta$. Unfortunately history went differently, so let
me explain what we have failed to do so far.

The geometry of $(\mathbb{R}^4_\theta,\mathcal{H},\mathcal{D})$ cannot
be the geometry of a spectral triple \cite{Connes:1996gi}, because the
spectrum of the Dirac operator $\mathcal{D}$ is continuous. It rather
fits into the axioms of ``non-compact spectral triples''
\cite{Gracia-Bondia:2001ct}.  In this framework the dimension of
$(\mathbb{R}^4_\theta,\mathcal{H},\mathcal{D})$ equals zero, not
four\footnote{The spectral triple for the noncommutative 4-torus has
  dimension four, not zero! The Hochschild dimension of
  $\mathbb{R}^4_\theta$ drops down to zero as well
  \cite{Connes:2001ef}.}, because $f \in \mathcal{S}(\mathbb{R}^4)$ is
trace-class so that $f|\mathcal{D}|^{-n}$ has vanishing Dixmier trace.
Thus, it is the requirement of rapid decay at infinity which brings
the dimension down to zero. Taking the $\star_\omega$ product for a
suitable class of analytic functions, we can keep the spectral
dimension at four. This is related to the notion of star triples in
\cite{Gracia-Bondia:2001ct}. 

The geometry is extracted from a spectral triple via 
\emph{states}---linear functionals $\chi
:\mathbb{R}^4_\theta \to \mathbb{C}$ such that\footnote{The 1 in
  $\chi(1)$ is thought to be the limit of a sequence of appropriate
  elements of $\mathbb{R}^4_\theta$.} $\chi(1)=1$ and $\chi(a^*\star
a) \geq 0$ for all $a \in \mathbb{R}^4_\theta$. We can view such a
state as an element of the multiplier algebra $\mathcal{M}_\theta$
through the formula
\begin{align}
\chi(a) &= \int d^4x \,\chi(x) a(x)\;, &
\int  d^4x \,\chi(x)&=1\;,&
\int  d^4x \,\big(\chi\star a^* \star a\big)(x) &\geq 0\;.
\end{align}
The space of states is made to a metric space by means of Connes'
distance formula
\begin{align}
\mathrm{dist}(\chi_1,\chi_2)=\sup_{a \in \mathcal{A}_\theta} 
\Big\{ \big|\chi_1(a)-\chi_2(a)\big|~:\quad
\big\|[\mathcal{D},a]\big\|_{\mathcal{B}(\mathcal{H})} \leq 1\Big\}\;.   
\end{align}
In the commutative case, for the states labelled by points according
to $\chi_y(x)=\delta^4(x-y)$, this formula returns the geodesic
distance of the points. For the standard model one recovers a discrete
Kaluza-Klein geometry in five dimensions \cite{Martinetti:2001fq}.

The states on commutative space suggest immediately to try whether
$\chi_y(x)=\delta^4(x-y)$ are states on $\mathbb{R}^4_\theta$ as well.
The answer is no\footnote{As a consequence, $\mathbb{R}^d_\theta$ is
  not the algebra of functions on some manifold.}. According to
Appendix B (which is copied from \cite{Gracia-Bondia:1987kw}) for the
two-dimensional case, there are functions $a \in \mathbb{R}^2_\theta$
and points $x\in \mathbb{R}^2$ such that $(a^*\star a)(x) <0$.
Moreover, the algebra $\mathbb{R}^2_\theta$ (functions of rapid decay
at infinity) is identified with the algebra of matrices of infinite
size.  The consequence is that field theory on $\mathbb{R}^4_\theta$
is rather a matrix theory than a traditional field theory on Euclidian
space, see \cite{Gonzalez-Arroyo:1983ac}. I would like to mention that
the matricial basis was crucial for Langmann's class of exactly
solvable quantum field theories in odd dimensions
\cite{Langmann:2002ai}.

Moreover, Figure~3 in Appendix A tells us that the product of two fields,
both of which with support in the interior of the area $\theta$, is,
to a high accuracy, zero. All this means that $\mathbb{R}^2_\theta$ is
divided into cells of area $\theta$, and fields on
$\mathbb{R}^2_\theta$ are characterised by assigning to each cell a
value. This picture is easily generalised to $\mathbb{R}^4_\theta$.
Now, since the interaction of fields on $\mathbb{R}^4_\theta$ is
smeared over the cell of size $|\det \theta|^{\frac{1}{2}}$, one would
expect that a quantum field theory on $\mathbb{R}^4_\theta$ is free
of divergences \cite{Cho:1999sg}. Performing the calculation of
Feynman graphs, however, one does encounter divergences, and these are
\emph{worse} than on commutative space-time. See sec.~\ref{pert}. 

How can we understand this puzzle? The Feynman rules are nothing but
the perturbative evaluation of the path integral (\ref{pathintegral}).
It seems that this kind of evaluation of (\ref{pathintegral}) somehow
brings the dimension of $\mathbb{R}^d_\theta$ back to $d$. The crucial
question here is the definition of the measure $\mathcal{D}[\phi]$ in
the path integral. The idea is to integrate over all possible
fields on $\mathbb{R}^d_\theta$. This is most conveniently done by
taking a basis, for instance the matricial basis $f_{mn}$ of Appendix
B in the two-dimensional case:
\begin{align}
\phi(x)=\sum_{m,n=0}^\infty \phi_{mn} f_{mn}(x)\;.
\end{align}
Thus, the measure should be $\mathcal{D}[\phi]=
\prod_{m,n=0}^\infty d\phi_{mn}$.  Next we have to specify 
the domain of integration. The Feynman rules correspond to integrating
\emph{all} $\phi_{mn}$ from $-\infty$ to $\infty$. Obviously this is
not compatible with the requirement that the $\{\phi_{mn}\}$ represent
an element of $\mathbb{R}^2_\theta$, which imposes
\cite{Gracia-Bondia:1987kw}
\begin{align}
\sum_{m,n=0}^\infty \big( (2m{+}1)^{2k}(2n{+}1)^{2k}\,|\phi_{mn}|^2
\big)^{\frac{1}{2}} <\infty \qquad \text{for all } k\;.
\label{phimn}
\end{align}
Integrating all $\phi_{mn}$ from $-\infty$ to $\infty$ we actually
include functions with unrestricted behaviour at infinity, and those
functions lead to a spectral dimension bigger than zero. In other
words, the discussion of the geometry of $\mathbb{R}^4_\theta$ tells
us that the usual Feynman rules are not adequate\footnote{The same
  criticism applies to canonical quantisation because the amplitudes
  cannot be promoted to harmonic oscillators at quantum numbers
  approaching infinity.} to quantum field theory on
$\mathbb{R}^4_\theta$. It is neither surprising nor a
problem that the standard Feynman graph approach to quantum field
theories on $\mathbb{R}^4_\theta$ fails miserably (see also the next
sections).

A possibility to stay within the class of fields of rapid decay at
infinity is to introduce a cut-off in the path integral measure, e.g.\ 
$\mathcal{D}[\phi]=\prod_{m,n=0}^L d\phi_{mn}$. A different view of
that cut-off is to regard all modes $\phi_{mn}$ as included, but with
the integral performed over the interval $[0,0]$ instead of
$[-\infty,\infty]$ if $m>L$ or $n>L$. This prescription can then be
deformed into a smooth cut-off with all $\phi_{mn}$ included, but with
$|\phi_{mn}|$ being of rapid decay at infinity. In this way we
integrate indeed over fields on $\mathbb{R}^d_\theta$. The cut-off can
be arbitrarily large and the cut-off function arbitrarily chosen; we
stay within the allowed class of fields as long as there is a cut-off
somewhere when approaching infinity.

Remarkably, this smooth cut-off version of the path integral is
exactly the way one proceeds in the exact renormalisation group
approach to renormalisation \cite{Wilson:1973jj}, see also
\cite{Polchinski:1983gv}. We see thus that the smooth cut-off in the
exact renormalisation group is not just a convenient trick to compute
the path integral, it is the direct consequence of the
zero-dimensional geometry\footnote{In that sense, the renormalisation
  group approach for commutative field theories is linked to fields of
  rapid decay in momentum space. It seems to be a degeneracy of
  commutative geoemtry that this restriction leads to the same results
  as the Feynman graph appraoch.} of $\mathbb{R}^d_\theta$. The
philosophy is then to redefine the theory in such a way that---once
specifying normalisation conditions---everything becomes independent
of the cut-off function and the value of the cut-off.  There are first
results \cite{Griguolo:2001ez} that at least scalar field theories on
$\mathbb{R}^4_\theta$ are renormalisable within the exact
renormalisation group approach.

\section{The Feynman graph approach to quantum field theories 
on $\mathbb{R}^4_\theta$}

We have argued in the last section that the Feynman graph approach to
quantum field theories on $\mathbb{R}^4_\theta$ does not correctly
reflect the geometry of $\mathbb{R}^4_\theta$. There has been,
however, an enormous amount of work along this line, which deserves a
few comments. We need a notation to distinguish the Feynman graph
approach from the true zero-dimensional $\mathbb{R}^4_\theta$. Let us
call the four-dimensional space where Feynman graph computations are
performed $\mathbb{R}^4_{nc}$.

The first contribution was the one-loop investigation of U(1)
Yang-Mills theory on $\mathbb{R}^4_{nc}$ by Mart\'{\i}n and
S\'anchez-Ruiz \cite{Martin:1999aq}. They found that all one-loop pole
terms of this model in dimensional regularisation\footnote{There is of
  course a problem extending $\theta$ to complex dimensions, this is
  however discussed in \cite{Martin:1999aq}.} can be removed by
multiplicative renormalisation (minimal subtraction) in a way
preserving the BRST symmetry. This is completely analogous to the
situation on the noncommutative 4-torus \cite{Krajewski:1999ja}.
Shortly later there appeared also an investigation of super-Yang-Mills
theory on $C^\infty(\mathbb{R}) \times \mathbb{T}^2_\theta$
\cite{Sheikh-Jabbari:1999iw}. In the following two years similar
one-loop calculations were performed for the
$\mathbb{R}^d_{nc}$-analogue of any existing commutative model.

The reason why these models became so attractive was the completely
unexpected discovery of quadratic infrared-like divergences, first in
quantum field theories of scalar fields \cite{Minwalla:1999px}, which
ruled out a perturbative renormalisation at higher loop order. At that
time it was an open question whether this is an artifact of scalar
fields or really a general feature. We have shown in
\cite{Grosse:2000yy} using power-law estimations for Bessel functions
that the sub-sector of Yang-Mills theory on $\mathbb{R}^4_{nc}$
given by repeated one-loop ghost propagator self-insertions is
renormalisable to any loop order. Shortly later it was demonstrated,
however, that there are one-loop Green's functions in Yang-Mills
theory on $\mathbb{R}^4_{nc}$ which have quadratic and linear
infrared-like divergences \cite{Matusis:2000jf}, which prevent any
renormalisation beyond one-loop.

In my opinion the most valuable contribution to field theories on
$\mathbb{R}^d_{nc}$ are the two articles \cite{Chepelev:1999tt,
  Chepelev:2000hm} by Chepelev and Roiban, in which they investigated
the convergence behaviour of massive quantum field theories \emph{at
  any loop order}. The essential technique is the representation of
Feynman graphs as ribbon graphs, drawn on an (oriented) Riemann
surface with boundary, to which the external legs of the graph are
attached. There are two important qualifiers for such a ribbon
graph, the index and the cycle number. The index is declared to
be one if the external lines attach to boundary components ``inside''
and ``outside'' of the graph, otherwise zero. The cycle number is the
number of homologically non-trivial cycles of the Riemann surface of
the total graph wrapped by the (sub)graph. Using this language and
sophisticated tricks for the manipulation of determinants, Chepelev
and Roiban were able to prove that in order to have convergence of the
integral, each subgraph must have one of the following properties:
\begin{enumerate}
\item The index is one and the external momenta are non-exceptional.
  
\item If the index is zero or the index is one but the external
  momenta are non-exceptional, then the power-counting degree of
  divergence of the graph is smaller than the dimension $d$ times the
  number of cycles.
\end{enumerate}
Thus, noncommutativity (index one and presence of cycles due to
non-planarity) improves the convergence of the integral. Integrals
associated to planar sectors are to be renormalised as in commutative
quantum field theories, they are not a problem. One has to make sure
however, that there are no divergences in non-planar sectors. It
turned out that there are two dangerous classes of non-planar
divergences, which in \cite{Chepelev:2000hm} are called ``Rings'' and
``Com''. Rings consists of a chain of divergent graphs stacked onto
the same cycle, they induce the problem first observed in
\cite{Minwalla:1999px}. Com's are index-one graphs with exceptional
external momenta due to momentum conservation, they correspond to
non-local divergences of the type $(\int \phi \star \phi)^2$. In
massless models they are catastrophic. Unfortunately this problem
seems to be completely ignored in literature.

\section{The non-locality of the divergences}
\label{pert}

All this is well-known by now and can be looked up in the literature,
nevertheless I would like to demonstrate the problem with quantum
field theories on $\mathbb{R}^4_{nc}$ by computing the ghost loop
contribution to the one-loop gluon two-point function. The necessary
Feynman rules adapted to the (Euclidian) BPHZL renormalisation scheme
\cite{Lowenstein:1975ps} are given by
\begin{fmffile}{fmfhesselberg-02}
\begin{align}
\parbox{33mm}{\begin{picture}(30,10)
\put(0,0){\begin{fmfgraph}(30,10)
\fmfleft{l}
\fmfright{r}
\fmf{ghost}{i,l}
\fmf{ghost}{r,i}
\end{fmfgraph}}
\put(6,7){\mbox{$p$}}
\put(20,7){\mbox{$-q$}}
\end{picture}} &= -\dfrac{1}{p^2+(s{-}1)^2 M^2 } \;,
\label{propagator}
\\
\parbox{33mm}{\begin{picture}(30,20)
\put(0,0){\begin{fmfgraph}(30,20)
\fmftop{t}
\fmfbottom{b1,b2}
\fmf{photon,tension=1.3}{t,i}
\fmf{ghost}{b2,i}
\fmf{ghost}{i,b1}
\fmffreeze
\fmf{phantom_arrow}{t,i}
\end{fmfgraph}}
\put(17,15){\mbox{$q,\nu$}}
\put(3,8){\mbox{$-p$}}
\put(25,8){\mbox{$r$}}
\end{picture}} &= 2 \mathrm{i} \, p^\nu \sin (\tfrac{1}{2} 
\theta^{\alpha\beta} q_\alpha r_\beta) \;,
\label{vertex}
\end{align}
for ghost propagator and ghost-gluon vertex, respectively. One has
$p{+}q=0$ in (\ref{propagator}) and $p{+}q{+}r=0$ in (\ref{vertex}).
We then compute the graph
\begin{align}
\gamma^{\mu\nu}(p,s) &= \parbox{43mm}{\begin{picture}(40,25)
\put(0,0){\begin{fmfgraph}(40,25)
\fmfleft{l}
\fmfright{r}
\fmf{photon,tension=2}{l,i1}
\fmf{phantom_arrow}{l,i1}
\fmf{photon,tension=2}{r,i2}
\fmf{phantom_arrow}{r,i2}
\fmf{ghost,left}{i1,i2}
\fmf{ghost,left}{i2,i1}
\end{fmfgraph}}
\put(3,15){\mbox{$p,\mu$}}
\put(32,15){\mbox{$-p,\nu$}}
\put(21,0){\mbox{$k{-}p$}}
\put(20,23){\mbox{$k$}}
\end{picture}}
\nonumber
\\*
& = \hbar \int \!\! \frac{d^4k}{(2\pi)^4} \,
\frac{k^\mu  (k{-}p)^\nu
\Big\{2 
- \mathrm{e}^{\mathrm{i} \theta^{\alpha\beta} p_\alpha k_\beta}
- \mathrm{e}^{-\mathrm{i} \theta^{\alpha\beta} p_\alpha k_\beta}
\Big\}
}{(k^2+(s{-}1)^2 M^2 )
((k{-}p)^2+(s{-}1)^2 M^2 )} \;.
\label{gamma}
\end{align}
\end{fmffile}%
The integral as it stands in (\ref{gamma}) is meaningless. We have to
define a renormalisation scheme which assigns to the graph in
(\ref{gamma}) a meaningful integral. Here one has to distinguish
between the planar part corresponding to the factor 2 in in $\{~\}$
and the non-planar part corresponding to the phase factors in $\{~\}$.
Let us first look at the planar part. The integral is quadratically
divergent, and according to the BPHZL scheme we replace the integrand
$I^{\mu\nu}(k;p,s)$ by the Taylor subtracted integrand
\begin{align}
(1-R) [I^{\mu\nu}(k;p,s)]\Big|_{s=1} := (1-t^1_{p,s-1}) (1-t^2_{p,s})
[I^{\mu\nu}(k;p,s)]\Big|_{s=1}\;,
\end{align}
where $t^\omega_{p,s'}[I]$ is the Taylor expansion of $I$ about $p=0$
and $s'=0$ up to total degree $\omega$. In the example (\ref{gamma}) we
have for the integrand without the factor 2 in $\{~\}$
\begin{align}
R[I^{\mu\nu}(k;p,s)] &= \frac{k^\mu k^\nu}{
(k^2)^2} 
+p_\rho \Big(
-\frac{k^\mu g^{\nu\rho}}{(k^2)^2}
+ \frac{2 k^\mu k^\nu k^\rho}{(k^2)^3}\Big) 
\nonumber
\\
&
+ p_\rho p_\sigma \Big( - \frac{2 k^\mu k^\sigma g^{\nu\rho} 
}{ (k^2+ M^2)^3}
+ \frac{4 k^\mu k^\nu k^\rho k^\sigma}{(k^2+ M^2)^4}\Big)
\nonumber
\\
& + (s{-}1)^2
\Big(\frac{2 M^2 k^\mu k^\nu}{(k^2+ M^2)^3}
+\frac{12 M^4 k^\mu k^\nu}{(k^2+ M^2)^4} \Big)
\nonumber
\\
& 
+ p_\rho (s{-}1) \Big(
\frac{4 M^2 k^\mu g^{\nu\rho}}{(k^2+ M^2)^3}
-  \frac{12 M^2 k^\mu k^\nu k^\rho}{(k^2+ M^2)^4}
\Big)\;.
\end{align}
Passing to $s=1$ and omitting the integrand which is odd under $k\to
-k$, we now get for the
planar part in (\ref{gamma})
\begin{align}
\gamma^{\mu\nu}_{planar,ren}(p,M) &=
2\hbar \int \!\! \frac{d^4k}{(2\pi)^4} \,
\Big(\frac{k^\mu  (k{-}p)^\nu}{k^2(k{-}p)^2} 
-\frac{k^\mu k^\nu}{(k^2)^2} 
\nonumber
\\
& \hspace*{3em} 
- p_\rho p_\sigma \Big( - \frac{2 k^\mu k^\sigma g^{\nu\rho} 
}{ (k^2+ M^2)^3}
+ \frac{4 k^\mu k^\nu k^\rho k^\sigma}{(k^2+ M^2)^4}\Big)\Big)\;.
\label{gammaren}
\end{align}
The integral (\ref{gammaren}) is absolutely convergent, see
\cite{Lowenstein:1975ps}.

Let us now compute the difference between
$\gamma^{\mu\nu}_{planar}$ and $\gamma^{\mu\nu}_{planar,ren}$ in
position space:
\begin{align}
&
\int \frac{d^4p}{(2\pi)^4}\Big( 
\gamma^{\mu\nu}_{planar}(p)-\gamma^{\mu\nu}_{planar,ren}(p,M)
\Big) \,\mathrm{e}^{-\mathrm{i} p_\lambda (x-y)^\lambda}
\nonumber
\\*
&= 2\hbar \,\delta^4(x-y) \,\int\!\! \frac{d^4k}{(2\pi)^4} \,
\frac{k^\mu k^\nu}{(k^2)^2} 
\nonumber
\\*
& + 2 \hbar \,\frac{\partial^2 \delta^4(x-y)}{\partial
  x^\rho \partial x^\sigma}\,\int\!\! \frac{d^4k}{(2\pi)^4} \,
\Big( \frac{2k^\mu k^\sigma g^{\nu\rho} 
}{ (k^2+ M^2)^3}
- \frac{4 k^\mu k^\nu k^\rho k^\sigma}{(k^2+ M^2)^4}\Big)\;.
\label{diff}
\end{align}
The result is zero unless $x=y$. We recall now that the Fourier
transformation of (\ref{gammaren}) is the ghost loop contribution to
the gluon two-point correlation function $\big\langle A^\mu(x)
A^\nu(y)\big\rangle$. In other words, replacing the meaningless
integral (\ref{gamma}) by the renormalised one (\ref{gammaren}), we
have only redefined (in fact correctly defined) the product of the
distributions $A^\mu(x)$ and $A^\nu(y)$ at coinciding points. This is
precisely the freedom which one has in a local quantum field theory
\cite{Epstein:gw}.

But what about the non-planar part? Although not being absolutely
convergent, the oscillating phase (see also Figure~3 in Appendix~A)
renders the integral actually convergent---provided that $p\neq 0$.
Thus the first possibility is to keep the non-planar part untouched in
the renormalisation scheme. But now there is a problem for $p\to 0$.
Note that the original (ill-defined) integral (\ref{gamma}) had no
problem at all for $p\to 0$, in fact the integral was zero for $p=0$.
But since we removed from the planar part its first Taylor
coefficients about $p=0$ in order to render the planar part integrable
for $k\to \infty$, the singular behaviour of the non-planar part for
$p\to 0$ is no longer compensated. For the one-loop graph it is not a
terrible problem\footnote{As long as one is not interested in
  producing numbers to be compared with experiments!}, but inserting
this result declared as finite as a subgraph into a bigger divergent
graph, the singular behaviour at $p\to 0$ makes the bigger graph
non-integrable.  We therefore find a fake infrared divergence, which
is only due to our (obviously wrong) renormalisation prescription which
treated the planar and non-planar parts \emph{differently}. This is
the so-called UV/IR-mixing, a name which is not very appropriate.

Since the above treatment of the non-planar part was unsuccessful, let
us also remove the first Taylor coefficients about $p=0$ from the
non-planar part. This Taylor expansion must not be applied to the
momenta in the phases, because the result would be an even worser
divergence in $k$ and not a milder one. The only possibility is to
define the renormalised total graph as
\begin{align}
\gamma^{\mu\nu}_{ren}(p,M) &=
\hbar \int \!\! \frac{d^4k}{(2\pi)^4} \,
\Big\{2 
- \mathrm{e}^{\mathrm{i} \theta^{\alpha\beta} p_\alpha k_\beta}
- \mathrm{e}^{-\mathrm{i} \theta^{\alpha\beta} p_\alpha k_\beta}
\Big\}
\Big(\frac{k^\mu  (k{-}p)^\nu}{k^2(k{-}p)^2} 
-\frac{k^\mu k^\nu}{(k^2)^2} 
\nonumber
\\*
& \hspace*{6em} 
- p_\rho p_\sigma \Big( - \frac{2 g^{\nu\rho} k^\mu k^\sigma 
}{ (k^2+ M^2)^3}
+ \frac{4 k^\mu k^\nu k^\rho k^\sigma}{(k^2+ M^2)^4}\Big)\Big)
\;.
\label{ren}
\end{align}
Now the integral converges absolutely, in particular
there is no problem any more for $p\to 0$. We have to verify, however,
that the change from $\gamma^{\mu\nu}(p)$ to
$\gamma^{\mu\nu}_{ren}(p,M)$ is compatible with locality. In the
planar part this change amounts to a redefinition of the product of
distributions at coinciding points. Let us thus evaluate the change in
the non-planar part, again in position space:
\begin{align}
&
\int \frac{d^4p}{(2\pi)^4}\Big( 
\gamma^{\mu\nu}_{non-planar}(p)-\gamma^{\mu\nu}_{non-planar,ren}(p,M)
\Big) \,\mathrm{e}^{-\mathrm{i} p_\lambda (x-y)^\lambda}
\nonumber
\\*
&= -\hbar \int \!\! \frac{d^4k}{(2\pi)^4} 
\int \!\! \frac{d^4p}{(2\pi)^4} \,
\Big(
\frac{k^\mu k^\nu}{(k^2)^2} 
- p_\rho p_\sigma \Big( \frac{2 g^{\nu\rho} k^\mu k^\sigma }{ 
(k^2+ M^2)^3}
- \frac{4 k^\mu k^\nu k^\rho k^\sigma}{(k^2+ M^2)^4}\Big)\Big)
\nonumber
\\*
& \hspace*{10em} \times \Big\{
\mathrm{e}^{-\mathrm{i} p_\alpha ((x-y)^\alpha
+\theta^{\alpha\beta} k_\beta)}
+\mathrm{e}^{-\mathrm{i} p_\alpha ((x-y)^\alpha
-\theta^{\alpha\beta} k_\beta)}
\Big\}
\nonumber
\\
&= - \frac{2\hbar}{(2\pi)^4 \,\det \theta} \bigg(
\frac{(\theta^{-1}{\cdot} (x{-}y))^\mu
(\theta^{-1}{\cdot} (x{-}y))^\nu}{
((\theta^{-1} {\cdot} (x{-}y))^2)^2}
+ 2 
\frac{(\theta^{-1})^{\mu\sigma}(\theta^{-1})_{\sigma}^{~\nu}}{
((\theta^{-1} {\cdot} (x{-}y))^2+M^2)^3}
\nonumber
\\*
&\qquad + 4 \frac{(\theta^{-2}{\cdot}(x{-}y))^\mu
(\theta^{-2}{\cdot}(x{-}y))^\nu}{
((\theta^{-1} {\cdot} (x{-}y))^2+M^2)^4}
+ 4 
\frac{(\theta^{-1})_{\alpha\beta}(\theta^{-1})^{\alpha\beta}
(\theta^{-1}{\cdot}(x{-}y))^\mu
(\theta^{-1}{\cdot}(x{-}y))^\nu}{
((\theta^{-1} {\cdot} (x{-}y))^2+M^2)^4}
\nonumber
\\*
&\qquad
- 8\frac{(\theta^{-3}{\cdot}(x{-}y))^\mu
(\theta^{-1}{\cdot}(x{-}y))^\nu}{
((\theta^{-1} {\cdot} (x{-}y))^2+M^2)^4}
- 8\frac{(\theta^{-1}{\cdot}(x{-}y))^\mu
(\theta^{-3}{\cdot}(x{-}y))^\nu}{
((\theta^{-1} {\cdot} (x{-}y))^2+M^2)^4}
\nonumber
\\*
&\qquad
+ 32\frac{
(\theta^{-1}{\cdot}(x{-}y))^\mu
(\theta^{-1}{\cdot}(x{-}y))^\nu
\;(\theta^{-2}{\cdot}(x{-}y))^\rho
(\theta^{-2}{\cdot}(x{-}y))_\rho}{
((\theta^{-1} {\cdot} (x{-}y))^2+M^2)^5}
\bigg).
\label{nonlocal}
\end{align}
There are contributions for $x\neq y$ in the non-planar part. In other
words, we have changed the non-planar part in a \emph{non-local} way
in order to achieve absolute convergence. This is not allowed in a
local quantum field theory, which means that our model on
$\mathbb{R}^4_{nc}$ is not renormalisable in the framework of local
quantum field theories.  

On the other hand, the result (\ref{nonlocal}) is exactly what one
should expect for a quantum field theory on $\mathbb{R}^4_\theta$:
Since the physical information cannot be localised at individual
points it must now be allowed to modify the product of distributions
not only at coinciding points but for the whole extended region of
volume $|\det \theta|^{\frac{1}{2}}$ in which information can be
concentrated. Unfortunately, this idea is not very well implemented in
the above calculation. The subtraction term is too much localised in
the planar part is not enough localised in the non-planar part. In my
opinion, the origin of this problem is the wrong choice of the measure
in the path integral which is used to derive the Feynman rules, see
sec~\ref{spec}.

\section{$\theta$-expanded field theories: general remarks}
\label{SWgen}

Let us now come to quantum field theories based on the Taylor-expanded
$\star_\omega$-product (\ref{staromega}) regarded order by order in
$\theta$. The philosophy here is to consider the Taylor expansion
$\star_\omega$ up to some finite order in $\theta$ only. In this way
we obtain a \emph{local} field theory on ordinary Euclidian or
Minkowski space for which standard Feynman graph techniques can safely
be applied. The only novelty is the presence of external fields
$\theta^{\alpha\beta}$ of power-counting dimension $-2$ which couple
to the commutative fields via partial derivatives. When restricting
the product
\begin{align}
(\phi_1 \star_\omega \phi_2)(x) &= \phi_1(x)\,\phi_2(x) 
\nonumber
\\
&+ \frac{\mathrm{i}}{2} \theta^{\alpha\beta}
\partial_\alpha \phi_1(x)\, \partial_\beta \phi_2(x)
- \frac{1}{8} \theta^{\alpha\beta}\theta^{\gamma\delta}
\partial _\alpha \partial_\delta \phi_1(x)\, 
\partial_\beta \partial_\gamma \phi_2(x) + \dots 
\end{align}
to some finite order, nothing is noncommutative, the second term on
the r.h.s.\ can equally well be written as $\frac{\mathrm{i}}{2}
\theta^{\alpha\beta} \partial_\beta \phi_2(x)\,\partial_\alpha
\phi_1(x)$.

The most interesting field theories are gauge theories\footnote{To the
  best of our knowledge, there are no fundamental scalar fields in
  nature---remember that the Higgs field is a noncommutative gauge
  field, and that supersymmetry is not found so far.}. The prototype
is Maxwell theory, the action functional of which, written in terms of
the $\star_\omega$-product, reads
\begin{align}
\Gamma[A] &=\int d^4x \,\Big(-\frac{1}{4 g^2} 
F_{\mu\nu}(x) \star_\omega F^{\mu\nu}(x)\Big)\;,
\label{Maxwell}
\\
F_{\mu\nu} &= \partial_\mu A_\nu-\partial_\mu A_\nu-\mathrm{i} A_\mu
\star_\omega A_\nu + \mathrm{i} A_\nu \star_\omega A_\mu 
\\
&= \partial_\mu A_\nu-\partial_\mu A_\nu + \theta^{\alpha\beta}
\partial_\alpha A_\mu\,\partial_\beta A_\nu 
-\frac{1}{24} \theta^{\alpha\beta} \theta^{\gamma\delta}
\theta^{\epsilon\zeta}
\partial_\alpha \partial_\gamma \partial_\epsilon A_\mu\,
\partial_\beta \partial_\delta \partial_\zeta A_\nu+ \dots\;.
\nonumber
\end{align}
Now, (\ref{Maxwell}) is an action functional for 
commutative boring photons, which is invariant under the infinitesimal
gauge transformation
\begin{align}
A_\mu &\mapsto A_\mu + (\partial_\mu \lambda - \mathrm{i}
A_\mu \star_\omega \lambda + \mathrm{i} \lambda \star_\omega A_\mu )
\label{gt}
\\
&= A_\mu + \partial_\mu \lambda 
+ \theta^{\alpha\beta}
\partial_\alpha A_\mu\,\partial_\beta \lambda
-\frac{1}{24} \theta^{\alpha\beta} \theta^{\gamma\delta}
\theta^{\epsilon\zeta}
\partial_\alpha \partial_\gamma \partial_\epsilon A_\mu\,
\partial_\beta \partial_\delta \partial_\zeta \lambda+ \dots\;.
\nonumber
\end{align}
But how is this possible, an action functional for photons which
transform in a very strange way? The answer was given by Seiberg and
Witten \cite{Seiberg:1999vs}: The photon is written in (\ref{Maxwell})
and (\ref{gt}) only in an extremely inconvenient way. There is a
change of variables
\begin{align}
A_\mu &= A'_\mu -\frac{1}{2} \theta^{\alpha\beta} A_\alpha'
(2 \partial_\beta A_\mu' - \partial_\mu A_\beta') + \dots\;,
\nonumber
\\
\lambda &= \lambda' - \frac{1}{2} \theta^{\alpha\beta} A_\alpha'
\partial_\beta \lambda' +\dots\;,
\label{SW}
\end{align}
which brings (\ref{Maxwell}) and (\ref{gt}) into the more pleasant
form
\begin{align}
\Gamma[A'] &=\int d^4x \,\Big(-\frac{1}{4 g^2} 
F'_{\mu\nu}(x) F^{\prime \mu\nu}(x) 
\nonumber
\\*
&\qquad
- \frac{1}{2 g^2} \theta^{\alpha\beta} F'_{\alpha\mu}(x) F'_{\beta\nu}(x)
F^{\prime \mu\nu}(x) + \frac{1}{8 g^2} \theta^{\alpha\beta} 
F'_{\alpha\beta}(x) F'_{\mu\nu}(x) F^{\prime \mu\nu}(x) + \dots \Big)\;,
\nonumber
\\*
F'_{\mu\nu} &= \partial_\mu A'_\nu-\partial_\nu A'_\mu\;,\qquad
\text{$\Gamma[A']$~~invariant under } 
A'_\mu \mapsto A'_\mu+\partial_\mu \lambda'\;.
\label{Maxwell'}
\end{align}
The last line in (\ref{Maxwell'}) is exact in $\theta$, it looks much
more familiar. Actually Seiberg and Witten formulated their result
differently. They interpreted the transformation (\ref{SW}) leading
from (\ref{Maxwell}) to (\ref{Maxwell'}) as an equivalence between a
noncommutative gauge theory and a commutative gauge theory. Now there
is a puzzle. Namely, from the noncommutative geometrical background,
the noncommutative field theory is given by a spectral triple which
can never be expressed in the language of manifolds. How can there be
a map to a commutative field theory? The solution is simple, but it
took me a long time to understand it: The initial formulation
(\ref{Maxwell}) was already in the framework of commutative local
geometry, because already there the $\star_\omega$ product was
restricted to some finite order in $\theta$. The transformation
(\ref{SW}) is merely a \emph{convenient change of variables within the
  same commutative framework}. The (very difficult) limit
where the order of $\theta$ goes to infinity is not discussed in this
approach.

\section{Lorentz invariance and Seiberg-Witten differential equation}

One may ask whether the Taylor expansion (\ref{staromega}) leading
from the non-local $\star$-product to the local $\star_\omega$-product
up to finite order in $\theta$, applied to a truly noncommutative
action functional $\Gamma[\hat{A}]$, can produce the $\theta$-expanded
action functional in the Seiberg-Witten transformed form
(\ref{Maxwell'}) in a single stroke, i.e.\ without passing through
(\ref{Maxwell}). This is possible indeed, it has something to do with
symmetry transformations of the noncommutative theory.

There has been a lot of confusion concerning the question of Lorentz
invariance of field theories on $\mathbb{R}^4_\theta$. Once and for
all, symmetries in the noncommutative world are automorphisms of the
algebra \cite{Connes:1996gi}. The algebra $\mathbb{R}^4_\theta$ is
determined by $\theta$ and the question is how $\theta$ is
characterised. We follow \cite{Doplicher:tu} and agree that $\theta$
is characterised by the two Lorentz invariants
$\theta^{\mu\nu}\theta_{\mu\nu}$ and $\epsilon_{\mu\nu\rho\sigma}
\theta^{\mu\nu} \theta^{\rho\sigma}$ when discarding dilatation and by
the ratio of these two when including dilatation. The individual
components $\theta^{\mu\nu}$ (with respect to a given basis) do not
have a physical meaning. The algebra
is $\mathbb{R}^4_\theta$, not $\mathbb{R}^4_{\theta^{\mu\nu}}$.

Let us be more explicit. Infinitesimal field transformations are
implemented by Ward identity operators
\begin{align}
W = \sum_i \Big\langle \delta \hat{\Phi}_i[\hat{\Phi}_k],
\frac{\delta}{\delta \hat{\Phi}_i}\Big\rangle\;,
\label{WIO}
\end{align}
where the index $i$ labels the different sorts of fields, here denoted
$\hat{\Phi}_i$. The Ward identity operator (\ref{WIO}) acts on
(sufficiently regular) functionals $\Gamma[\hat{\Phi}_i]$ in a
derivational manner:
\begin{align}
W\Gamma[\hat{\Phi}_i] = \sum_j \Big\langle \delta \hat{\Phi}_j[\hat{\Phi}_k],
\frac{\delta \Gamma[\hat{\Phi}_i]}{\delta \hat{\Phi}_j}\Big\rangle
= \lim_{\epsilon \to 0} \frac{1}{\epsilon}\Big(
\Gamma[\hat{\Phi}_i+\epsilon \delta\hat{\Phi}_i[\hat{\Phi}_k]]
- \Gamma[\hat{\Phi}_i]\Big)\;.
\end{align}

We are interested in a set $\mathcal{S}$ of symmetry transformations
of the action functional, $W^I\Gamma=0$, $I \in \mathcal{S}$. This set
is required to be complete, $[W^{I},W^{I'}] = \sum_n W^{I_n}$, $I_n
\in \mathcal{S}$. In particular, we are interested in gauge
transformation $G$ and Lorentz transformation $L$ which satisfy
\begin{align}
[W^L,W^L] &\subset W^L\;, &
[W^G,W^G] &\subset W^G\;, &
[W^G,W^L] &\subset W^G\;.
\label{WW}
\end{align}
The Lorentz transformation has for the field $\hat{A}$ of Yang-Mills
theory on $\mathbb{R}^4_\theta$ the symbolic form
\begin{align}
W^L=   
\Big\langle \delta^L \hat{A} ,\frac{\delta}{\delta \hat{A}}\Big\rangle
+ \Big\langle \delta^L \theta ,\frac{\delta}{\delta \theta}\Big\rangle\;,
\label{WL}
\end{align}
it is a symmetry of the Yang-Mills action functional, and (\ref{WW})
is satisfied \cite{Bichl:2001yf}. It is essential that in (\ref{WL})
the sum of the $\hat{A}$ and the $\theta$-transformation appears, the
individual transformations do not have any meaning. Neither they are
symmetries of the action functional, nor they fulfil (\ref{WW}). But
if one really insists on transforming $\hat{A}$ only, then at least
this transformation $\tilde{W}^L_{\hat{A}}$, which cannot be a
symmetry of the action functional, must satisfy 
\begin{align}
[\tilde{W}^L_{\hat{A}}, W^G] \subset W^G\;.
\label{g-equiv}
\end{align}
The condition (\ref{g-equiv}) guarantees that $\tilde{W}^L_{\hat{A}}
\Gamma[\hat{A}] \neq 0$, which can be regarded as the particle Lorentz
symmetry breaking, is a gauge-invariant quantity \cite{Bichl:2001yf}.
Otherwise $\tilde{W}^L_{\hat{A}}$ is completely unphysical. It is then
somehow natural to make the ansatz
\begin{align}
W^L&= \tilde{W}^L_{\hat{A}} + \tilde{W}^L_{\theta} \;, 
\nonumber
\\*
\tilde{W}^L_{\hat{A}} &= 
\Big\langle \delta^L \hat{A} - \delta^L \theta \,\frac{d \hat{A}}{d
  \theta} ,\frac{\delta}{\delta \hat{A}}\Big\rangle\;, &
\tilde{W}^L_{\theta} &= 
\Big\langle \delta^L \theta ,\frac{\delta}{\delta \theta}\Big\rangle
+ \Big\langle \delta^L \theta \,\frac{d \hat{A}}{d
  \theta} ,\frac{\delta}{\delta \hat{A}}\Big\rangle\;,
\end{align}
where $\frac{d \hat{A}}{d\theta}$ is, for the time being, just a
symbol. The condition (\ref{g-equiv}) determines $\frac{d
  \hat{A}}{d\theta}[\hat{A}]$, which thus becomes a concrete (but not
unique) function of $\hat{A}$. The equation $\frac{d
  \hat{A}}{d\theta}=\frac{d \hat{A}}{d\theta}[\hat{A}]$ looks formally
like a differential equation---the Seiberg-Witten differential
equation. Now we can define the following Taylor expansion of the
action functional $\Gamma[\hat{A}]$:
\begin{align}
\Gamma^{(n)}[A] &:= \sum_{j=0}^n \frac{1}{j!} (\theta)^j 
\Big( \big(\tilde{W}^1_{\theta}\big)^j 
\Gamma[\hat{A}]\Big)_{\theta=0}\;, & 
\delta^1 \theta &:= 1\;,&
A &:= \big(\hat{A}\big)_{\theta=0}\;.
\label{th-Gamma}
\end{align}
By construction, the action functional $\Gamma^{(n)}[A]$ describes a
commutative Yang-Mills theory (coupled to the external field $\theta$)
which is invariant under commutative gauge and Lorentz transformations
at any cut-off order $n$ in $\theta$, see \cite{Bichl:2001yf}. We have
thus obtained (\ref{Maxwell'}) up to any desired order in a single
stroke.

\section{Quantisation of $\theta$-expanded field theories}
\label{thexp}

{}From a physical point of view, $\theta$-expanded quantum field
theories are not so interesting, because they are local and therefore
show all the the problems discussed in sec.~\ref{farewell}. They have
a very interesting structure, though, because the appearance of a
field $\theta$ of power-counting dimension $-2$ makes them
power-counting \emph{non-renormalisable}. It could seem, therefore,
that it is not very useful to study such a model as a quantum field
theory.  However, at the same time where $\theta$ leads to an
explosion of the number of divergences, it also provides the means to
absorb a considerable fraction of these divergences through field
redefinitions. A field redefinition is a non-linear generalisation of
the usual wave function renormalisation, a generalisation which is
possible precisely because there is a field of negative power-counting
dimension. And there could be symmetries in the
$\theta$-expanded action which would prevent the appearance of other
divergences. There is thus a race between the number of divergences
created by $\theta$ and the number of divergences absorbable by
(unphysical) field redefinitions or avoided by symmetries.

The winner is probably the creator of divergences, but this is a
conjecture only. In this case, although there is at any given order
$n$ in $\theta$ a finite number of new interaction terms only, the
theory looses all predictability in the limit $n\to \infty$. There are
however signs for hope. First, all superficial divergences in the
photon self-energy in $\theta$-expanded Maxwell theory are
field redefinitions, to all order $n$ in $\theta$ and any loop order
\cite{Bichl:2001cq}. For the photon self-energy the field
redefinitions win the race. 

A direct search for symmetries was not successful so far so that the
only chance to detect them is to perform some loop calculations. Due
to the extremely rich tensorial structure in presence of $\theta$,
these calculations are extremely difficult to perform, even for the
one-loop photon self-energy in $\theta$-deformed Maxwell theory to
second order in $\theta$ \cite{Bichl:2001nf}. The photon three-point
function which is of at least third order in $\theta$ is already
beyond the means.

The simplest model to study other Green's functions than the
self-energy is $\theta$-deformed QED. I have computed in
\cite{Wulkenhaar:2001sq} all divergent one-loop Green's functions up
to first order in $\theta$. The result was astonishing. Although not
renormalisable at the considered order, there was in the massless case
only a single divergence more than those absorbable by field
redefinitions, where four exceeding divergent terms were to expect. In
the massive case (where the mass term is inserted directly into the
Dirac action) things become really bad so that this work suggests that
fermion masses should be introduced via a Higgs mechanism.

The results of \cite{Wulkenhaar:2001sq} provide a very strong signal
that new symmetries in $\theta$-expanded field theories exist indeed.
Since the initial action functional comes via (\ref{th-Gamma}) from an
action functional on $\mathbb{R}^4_\theta$, it seems plausible that
these symmetries are already present in the truly noncommutative field
theory. For me this is the justification to study $\theta$-expanded
quantum field theories: Although being completely different from
quantum field theories on $\mathbb{R}^4_\theta$, the otherwise
unphysical $\theta$-expanded models may provide valuable information
about the symmetries of the really interesting noncommutative models.
My feeling is that these symmetries come through the spectral action.
The spectral action is invariant under all unitarities of the Hilbert
space, not only those coming from the algebra. The problem is to make
this idea explicit.

The loop calculations of \cite{Bichl:2001nf, Wulkenhaar:2001sq} were
performed for the $\theta$-expanded action which comes out of
(\ref{th-Gamma}), with the standard commutative gauge invariance
(\ref{Maxwell'}). As we have shown in \cite{Grimstrup:2002af}, very
similar computations are possible when starting directly from the
action functional for the $\star_\omega$-product, see (\ref{Maxwell}).
The only difference is that now the gauge symmetry is non-linearly
realised so that the whole machinery of external fields and
Slavnov-Taylor identities must be used. It is not sufficient to write
down the BRST transformations only. We looked as in
\cite{Wulkenhaar:2001sq} at $\theta$-expanded QED up to first order in
$\theta$, and to our great surprise we found---up to field
redefinitions---exactly the same result as in
\cite{Wulkenhaar:2001sq}. This seems to indicate that the
Seiberg-Witten map (\ref{SW}) is an unphysical change of variables
also on quantum level.

This is true to some extent, but there is a subtlety. One can perform
the change of variables before or after quantisation. Changing the
variables $\Phi'=\Phi'[\Phi]$ after quantization, i.e.\ performing a
change of the dummy integration variables in the path integral
(\ref{pathintegral}), one obtains exactly the same Green's functions.
This was to expect from the general equivalence theorem
\cite{Kamefuchi:sb}.  The changes in the Feynman rules from $\Phi'$ to
$\Phi$ are compensated by graphs involving the modified source term
$\langle J, \Phi'[\Phi]\rangle$. In principle one would also expect
contributions from field redefinition ghosts, but here the propagator
equals $1$ so that there is no contribution at least for certain
regularisation schemes. On the other hand, changing the variables in
the action functional before inserting it into the path integral, the
outcome is expected to be different. However, at first order in
$\theta$ only, the difference to the other method is a field
redefinition.

\section{Outlook}

Trial-and-error is the best method to start exploring a new world. We
have collected a big amount of empirical data on Feynman graph
computations of quantum field theories on noncommutative
$\mathbb{R}^4$. These theories are one-loop renormalisable and show at
higher loop order a new type of infrared-like non-local divergences.
Any model one can possibly think of has been studied. Everything is
covered by the power-counting theorem \cite{Chepelev:2000hm} (when
extended to the massless case \`a la Lowenstein). This is the most
rigorous result so far.  On the Taylor expanded side,
$\theta$-expanded field theories suggest that there are new
symmetries. Further going loop calculations are not possible in future
due to the enormous complexity of the outcome.  Thus, the
trial-and-error epoch has finished.

Now it is time for a more systematic approach. As argued in
sec.~\ref{spec}, the Feyman graph approach does not correctly reflect
the geometry of $\mathbb{R}^4_\theta$. Instead, one has to introduce a
smooth cut-off in the path integral and to compute it directly with
methods of the exact renormalisation group approach
\cite{Wilson:1973jj, Polchinski:1983gv}.

\begin{appendix}

\renewcommand{\theequation}{\Alph{section}.\arabic{equation}}
\makeatletter
\@addtoreset{equation}{section}
\makeatother
\renewcommand{\floatpagefraction}{1}
\renewcommand{\textfraction}{0}

\section{An example of the $\star$-product in two dimensions}

We consider the following function on $\mathbb{R}^2$
\begin{align}
f_{\vec{a},\vec{L}}(\vec{x}) &= \prod_{i=1}^2 f_{a_i,L_i}(x_i)\;, 
\nonumber
\\
f^N_{a_i,L_i}(x_i) &= \left\{ \begin{array}{cl} 
\cos (\frac{x_i-a_i}{L_i}) & \text{ for } 
a_i -\frac{(2N{+}1)\pi L_i}{2} \leq x_i \leq a_i +
\frac{(2N{+}1)\pi L_i}{2} 
\\[1ex]
0 & \text{ for } |x_i-a_i| > \frac{(2N{+}1)\pi L_i}{2} 
\end{array} \right.
\label{cos}
\end{align}
Clearly $f^N_{\vec{a},\vec{L}}(\vec{x}) \in \mathcal{S}(\mathbb{R}^2)$
(piecewise) for finite $N$ because for multi-indices
$\alpha=\{\alpha_i\}$ and $\beta=\{\beta_i\}$ one has $|(x)^\alpha
(\partial_x)^\beta f^N_{\vec{a},\vec{L}}(\vec{x})| \leq \prod_{i=1}^2
L_i^{-\beta_i} \big(|a_i|+\frac{(2N{+}1)\pi L_i}{2}\big)^{\alpha_i}$.
It is now an elementary calculation to compute the $\star$-product
(\ref{starprod}) of two functions (\ref{cos}):
\begin{align}
&\big(f^N_{\vec{a},\vec{L}} \star f^N_{\vec{b},\vec{L}'}\big)(\vec{x}) 
\label{coscos}
\\*
&=
\Big( \frac{1}{4} \frac{(\mathrm{-i})}{2\pi} 
\sum_{\epsilon,\epsilon'=\pm 1} 
\mathrm{e}^{-\mathrm{i}(\epsilon\frac{x_1-a_1}{L_1}
+ \epsilon'\frac{x_2-b_2}{L_2'} 
+\epsilon\epsilon'\frac{\theta}{2L_1 L_2'})}
\nonumber
\\*
& \hspace*{1em} \times \!\!\!\!
\sum_{\epsilon'',\epsilon'''=\pm 1} \!\!\!\!
\epsilon''\epsilon''' \,
\mathcal{G}\big[\tfrac{2L_1 L_2'}{\theta}
(\tfrac{x_1-a_1}{L_1}{+}\epsilon''' \tfrac{(2N{+}1)\pi}{2}
{+}\epsilon'\tfrac{\theta}{2 L_1 L_2'})
(\tfrac{x_2-b_2}{L_2'}{+}\epsilon'' \tfrac{(2N{+}1)\pi}{2} 
{+}\epsilon \tfrac{\theta}{2L_1 L_2'})\big]
\Big)
\nonumber
\\*
& 
\times 
\Big( \frac{1}{4} \frac{\mathrm{i}}{2\pi} 
\sum_{\epsilon,\epsilon'=\pm 1} 
\mathrm{e}^{\mathrm{i}(\epsilon\frac{x_2-a_2}{L_2}
+ \epsilon' \frac{x_1-b_1}{L_1'} 
+\epsilon\epsilon'\frac{\theta}{2L_2 L_1'})}
\nonumber
\\
& \hspace*{1em} \times \!\!\!\!
\sum_{\epsilon'',\epsilon'''=\pm 1} \!\!\!\!
\epsilon''\epsilon''' \,
\mathcal{G}\big[{-}\tfrac{2L_2 L_1'}{\theta}
(\tfrac{x_2-a_2}{L_2}{+}\epsilon''' \tfrac{(2N{+}1)\pi}{2}
{+}\epsilon'\tfrac{\theta}{2L_2 L_1'})
(\tfrac{x_1-b_1}{L_1'} {+}\epsilon'' \tfrac{(2N{+}1)\pi}{2}
{+}\epsilon \tfrac{\theta}{2L_2 L_1'})
\big]
\Big)\,,
\nonumber
\end{align}
where $\theta\equiv \theta^{12}=-\theta^{21}$,
\begin{align}
\mathcal{G}[u] &:= \sum_{n=1}^\infty \frac{(\mathrm{i} u)^n}{n \,n!} 
= \mathrm{ci}(u)-\gamma_E -\ln(u)+\mathrm{i}\,\mathrm{si}(u)\;,
\hspace*{-\textwidth}
\\
\mathrm{ci}(u)& =-\int_u^\infty \frac{dt\,\cos t}{t}
= \gamma_E +\ln(u)+\int_0^u \frac{dt\,(\cos t-1)}{t} \;, &
\mathrm{si}(u)& =\int_0^u \frac{dt\,\sin t}{t}\;, 
\nonumber
\end{align}
and $\gamma_E= 0.577216\dots$. In the limit $N\to \infty$ one recovers the
$\star_\omega$ product of the cosine functions:
\begin{align}
&
\big(f^\infty_{\vec{a},\vec{L}}\star_\omega 
f^\infty_{\vec{b},\vec{L}'}\big)(\vec{x})
\nonumber
\\*
&= \Big(\frac{1}{4} \sum_{\epsilon,\epsilon'=\pm 1} 
\mathrm{e}^{-\mathrm{i}(\epsilon\frac{x_1-a_1}{L_1}
+ \epsilon' \frac{x_2-b_2}{L_2'} + \epsilon\epsilon'
\frac{\theta}{2L_1 L_2'})}\Big)
\Big(\frac{1}{4} \sum_{\epsilon,\epsilon'=\pm 1} 
\mathrm{e}^{\mathrm{i}(\epsilon\frac{x_2-a_2}{L_2}
+ \epsilon' \frac{x_1-b_1}{L_1'} + \epsilon\epsilon'
\frac{\theta}{2L_2 L_1'})}\Big)
\nonumber
\\*
&= \Big(\cos \big(\tfrac{x_1-a_1}{L_1}\big)
\cos \big(\tfrac{x_2-a_2}{L_2}\big)\Big) \star_\omega 
\Big(\cos \big(\tfrac{x_1-b_1}{L_1'}\big)
\cos \big(\tfrac{x_2-b_2}{L_2'}\big)\Big) \;.
\label{coscos2}
\end{align}

It is illuminating to plot (\ref{coscos}) for various values of
$\theta$ and $N$. For simplicity we choose 
$a_i=b_i=0$ and $L_i=L_i'=L$. The result for $N\in\{0,1\}$
is shown in Figure~1 for $\theta=L^2$ and the cut with the plane 
$x_1=x_2$ for $\theta \in \{0.1L^2,L^2,3L^2\}$ in Figure~2.%
\begin{figure}[h]
\parbox{80mm}{\begin{picture}(80,45)
\put(0,-337){\epsfig{file=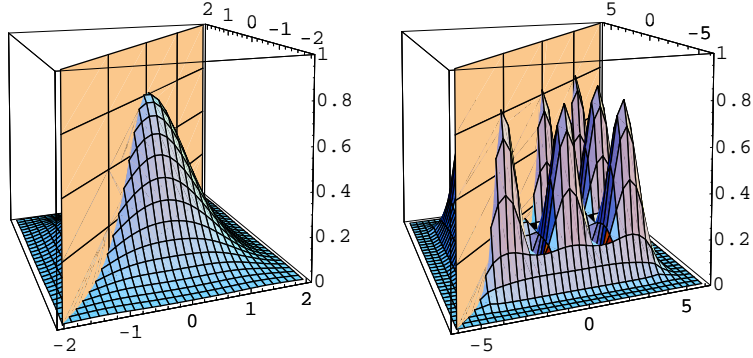,scale=1.3}}
\end{picture}}
\caption{The functions $f^0\star f^0$ (left) and 
  $f^1\star f^1$ (right) at $\theta=L^2$, where $f^N\equiv
  f^N_{(0,0),(L,L)}$. The $x_1,x_2$ axes are in units of $L$. The cut 
  with the plane $x_1=x_2$ is shown in Figure 2 for various values of
  $\theta$.}
\end{figure}
\begin{figure}[h]
\parbox{80mm}{\begin{picture}(80,37)
\put(-25,-493){\epsfig{file=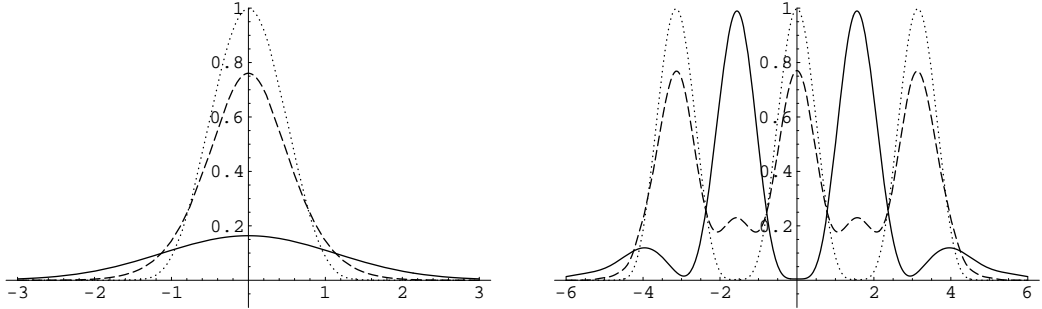,scale=1.8}}
\end{picture}}
\caption{The cut $x_1=x_2$ through the functions $f^0\star f^0$ (left) and 
  $f^1\star f^1$ (right) at $\theta=0.1L^2$ (dots), $\theta=L^2$
  (dashes) and $\theta=3L^2$ (solid), where
  $f^N\equiv f^N_{(0,0),(L,L)}$. The $\star$-product is \emph{smooth and
    non-local}. For $\theta\to0$ the commutative case is
  well approximated.}
\end{figure}
Actually the way one should read Figure 2 is the following. One should
regard $\theta$ as fixed and what varies is the characteristic length
$L$. For $L^2\gg \theta$ the influence of $\theta$ can be neglected,
and the $\star$-product agrees to
high precision with the usual commutative product of functions.  For
$L\ll\theta$ the situation is drastically different.  The
$\star$-product is distributed over a region of size $\sqrt{\theta}$,
whatever $L$ is, at the same time the amplitudes are damped. 
This is impressively shown in Figure~3,%
\begin{figure}[h]
\parbox{80mm}{\begin{picture}(80,42)
\put(-28,-520){\epsfig{file=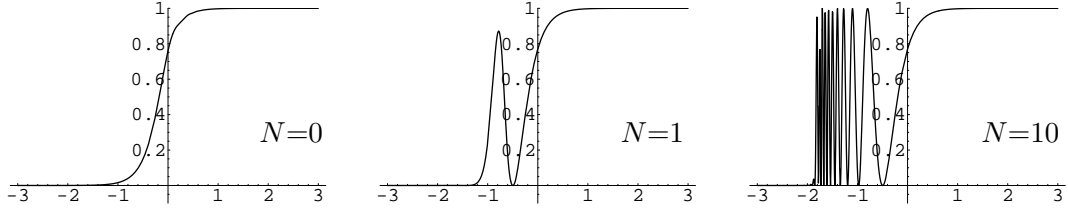,scale=1.85}}
\put(35,8){\mbox{\small$N{=}0$}}
\put(83,8){\mbox{\small$N{=}1$}}
\put(131,8){\mbox{\small$N{=}10$}}
\end{picture}}
\caption{The value $(f^N\star f^N)(0)$ 
  over $\log_{10}(\frac{L^2}{\theta})$ for $N\in\{0,1,10\}$, where
  $f^N\equiv f^N_{(0,0),(L,L)}$. This shows in a striking manner that the
  $\star$-product $\theta$ acts as a horizon. Oscillations of
  characteristic area smaller than $\theta$ are filtered out. }
\end{figure}
where the value of the product at $0$ is plotted over
$\log_{10}(L^2/\theta)$. 
If the functions are extremely localised, i.e.\ if
$\big(\frac{(2N+1)\pi L}{2}\big)^2 \ll \theta$, their product is zero
to a high precision. Thus, $\theta$ acts as a horizon: Oscillations
contained in an area smaller than $\theta$ are smoothed away.  
They do not carry any physical information. See also \cite{Bars:2001dy}.

\section{The matricial basis of $\mathbb{R}^2_\theta$}

The following is copied from \cite{Gracia-Bondia:1987kw}, adapted to
our notation. It proves that evaluation at $x \in \mathbb{R}^2$ is not
a state on $\mathbb{R}^2_\theta$.

The Gaussian 
\begin{align}
f_0(x) &= 2 \mathrm{e}^{-\frac{1}{\theta}(x_1^2+x_2^2)}\;,
\end{align}
with $\theta\equiv \theta^{12}=-\theta^{21}>0$, is an idempotent,
\begin{align}
(f_0 \star f_0)(x) &= 4 \int d^2y\int \frac{d^2k}{(2\pi)^2} \,
\mathrm{e}^{-\frac{1}{\theta}(2 x^2 +y^2+ 2x\cdot y + x \cdot \theta
  \cdot k +  \frac{1}{4} \theta^2 k^2) + \mathrm{i} k\cdot y}
\nonumber
\\
&= \frac{\theta}{\pi} \int d^2k \,
\mathrm{e}^{-\frac{1}{\theta}(x^2 + x \cdot \theta \cdot k 
+\mathrm{i} (k \cdot x) \theta
+  \frac{1}{2} \theta^2 k^2) }
= f_0(x)\;.
\label{ff}
\end{align}
We consider creation and annihilation operators
\begin{align}
a &= \frac{1}{\sqrt{2}}(x_1+\mathrm{i} x_2)\;, & 
\bar{a} &= \frac{1}{\sqrt{2}}(x_1-\mathrm{i} x_2)\;, 
\nonumber
\\*
\frac{\partial}{\partial a} &= \frac{1}{\sqrt{2}}(\partial_1 
- \mathrm{i} \partial_2 )\;, & 
\frac{\partial}{\partial \bar{a}} &= \frac{1}{\sqrt{2}}(\partial_1 
+ \mathrm{i} \partial_2 )\;.
\end{align}
For any $f \in \mathbb{R}^2_\theta$ we have
\begin{align}
(a \star f)(x) &= a(x) f(x) 
+ \frac{\theta}{2} \frac{\partial f}{\partial \bar{a}}(x)\;, &
(f \star a)(x) &= a(x) f(x) 
- \frac{\theta}{2} \frac{\partial f}{\partial \bar{a}}(x)\;, 
\nonumber
\\
(\bar{a} \star f)(x) &= \bar{a}(x) f(x) 
- \frac{\theta}{2} \frac{\partial f}{\partial a}(x)\;, &
(f \star \bar{a})(x) &= \bar{a}(x) f(x) 
+ \frac{\theta}{2} \frac{\partial f}{\partial a}(x)\;. 
\end{align}
This implies $\bar{a}^{\star m} \star f_0=2^m \bar{a}^m f_0$, 
$f_0 \star a^{\star n} =2^n a^n f_0$ and 
\begin{align}
a \star \bar{a}^{\star m} \star f_0 &= \left\{\begin{array}{cl}
m\theta (\bar{a}^{\star (m-1)} \star f_0) & \text{ for } m\geq 1
\\
0 & \text{ for } m =0
\end{array}\right.
\nonumber
\\
f_0 \star a^{\star n} \star \bar{a} &= \left\{\begin{array}{cl}
n \theta  (f_0 \star a^{\star (n-1)})\;\, & \text{ for } n\geq 1
\\
0 & \text{ for } n =0
\end{array}\right.
\label{faa}
\end{align}
where $a^{\star n} = a \star a \star \dots \star a$ ($n$
factors) and similarly for $\bar{a}^{\star m}$. Now, defining
\begin{align}
f_{mn} &:=\frac{1}{\sqrt{n! m! \,\theta^{m+n}}} \, 
\bar{a}^{\star m} \star f_0 \star a^{\star n}
\\
& =  \frac{1}{\sqrt{n! m! \,\theta^{m+n}}} \sum_{k=0}^{\min(m,n)} 
\binom{m}{k} \binom{n}{k} \,k! \,2^{m+n-2k}\, \theta^k \,\bar{a}^{m-k} 
\,a^{n-k} 
f_0\;,
\nonumber
\end{align}
(the second line is proved by induction) it follows from (\ref{faa})
and (\ref{ff}) that
\begin{align}
(f_{mn} \star f_{kl})(x) = \delta_{nk} f_{ml}(x)\;.  
\label{fprod}
\end{align}
The multiplication rule (\ref{fprod}) identifies the $\star$-product
with the ordinary matrix product:
\begin{align}
a(x) &= \sum_{m,n=0}^\infty a_{mn} f_{mn}(x)\;,& 
b(x) &= \sum_{m,n=0}^\infty b_{mn} f_{mn}(x) 
\nonumber
\\*
\Rightarrow\quad (a\star b)(x) &= \sum_{m,n=0}^\infty (ab)_{mn} 
f_{mn}(x)\;, &
(ab)_{mn} &= \sum_{k=0}^\infty a_{mk} b_{kn}\;.
\end{align}
In order to describe elements of $\mathbb{R}^2_\theta$ the sequences
$\{a_{mn}\}$ must be of rapid decay \cite{Gracia-Bondia:1987kw}:
\begin{align}
\sum_{m,n=0}^\infty a_{mn} f_{mn} \in \mathbb{R}^2_\theta \qquad
\text{iff} \quad \sum_{m,n=0}^\infty \big((2m{+}1)^{2k}(2n{+}1)^{2k}
|a_{mn}|^2\big)^{\frac{1}{2}} < \infty \quad \text{for all } k\;.
\end{align}
Finally, using (\ref{ff}) we compute
\begin{align}
\int d^2 x \, f_{mn}(x) &= 
\frac{1}{\sqrt{m! n!\, \theta^{m+n}}} \int d^2x\, \big(
\bar{a}^{\star m} \star f_0 \star f_0 \star a^{\star n} \big)(x)
\nonumber
\\*
&= \frac{1}{\sqrt{m! n!\, \theta^{m+n}}} \int d^2x\, \big(
f_0 \star a^{\star n} \star \bar{a}^{\star m} \star f_0 \big)(x)
\nonumber
\\*
&= \delta_{mn} \int d^2x f_0(x) = 2 \pi \theta \delta_{mn}\;.
\label{intfmn}
\end{align}

Now we return to the question of states. We clearly have
\begin{align}
(f_{mn}^* \star f_{mn})(x) &= (f_{nm}\star f_{mn})(x)=f_{nn}(x)\;,
\label{f*f}
\end{align}
and $f_{11}(x) = \frac{2}{\theta}\big(4x_1^2+4x_2^2 -\theta\big)
\mathrm{e}^{-\frac{1}{\theta}(x_1^2+x_2^2)}< 0$ for $4x_1^2+4x_2^2
<\theta$. Thus, $\delta$-distributions cannot be states on
$\mathbb{R}^d_\theta$. On the other hand, (\ref{f*f}) and (\ref{intfmn})
imply that $ \chi_n(x) = \frac{1}{2\pi\theta} f_{nn}(x) $ are states
on $\mathbb{R}^2_\theta$. The basis $f_{mn}$ was used in
\cite{Langmann:2002ai} to construct a new class of exactly solvable
quantum field theories.

\end{appendix}


\begin{thebibliography}{99}

\bibitem{Doplicher:tu} S.~Doplicher, K.~Fredenhagen and J.~E.~Roberts,
  ``The Quantum Structure Of Space-Time At The Planck Scale And
  Quantum Fields,'' Commun.\ Math.\ Phys.\ {\bf 172} (1995) 187.

\bibitem{Connes:tu} A.~Connes, ``Noncommutative Geometry And
  Reality,'' J.\ Math.\ Phys.\ {\bf 36} (1995) 6194.

\bibitem{Connes:1996gi} A.~Connes, ``Gravity coupled with matter and
  the foundation of non-commutative geometry,'' Commun.\ Math.\ Phys.\ 
  {\bf 182} (1996) 155 [arXiv:hep-th/9603053].

\bibitem{Connes} 
A.~Connes, ``Noncommutative geometry'', Academic Press, 1994.

\bibitem{Gracia-Bondia:tr} J.~M.~Gracia-Bond\'{\i}a, J.~C.~V\'arilly
  and H.~Figueroa, ``Elements Of Noncommutative Geometry,'' {\it
    Boston, USA: Birkhaeuser (2001) 685 p}.

\bibitem{Carminati:1997ej} L.~Carminati, B.~Iochum and T.~Sch\"ucker,
  ``Noncommutative Yang-Mills and noncommutative relativity: A bridge
  over trouble water,'' Eur.\ Phys.\ J.\ C {\bf 8} (1999) 697
  [arXiv:hep-th/9706105].

\bibitem{Chamseddine:1996zu}
A.~H.~Chamseddine and A.~Connes,
``The spectral action principle,''
Commun.\ Math.\ Phys.\  {\bf 186} (1997) 731
[arXiv:hep-th/9606001].

\bibitem{Kopf}
T. Kopf and M. Paschke, ``A Spectral Quadruple for de Sitter Space,''
arXiv:math-ph/0012012.

\bibitem{Strohmaier}
A. Strohmaier, ``On Noncommutative and semi-Riemannian Geometry,''
arXiv:math-ph/0110001.

\bibitem{Gracia-Bondia:2001ct}
J.~M.~Gracia-Bond\'{\i}a, F.~Lizzi, G.~Marmo and P.~Vitale,
``Infinitely many star products to play with,''
JHEP {\bf 0204} (2002) 026
[arXiv:hep-th/0112092].

\bibitem{Kamefuchi:sb}
S.~Kamefuchi, L.~O'Raifeartaigh and A.~Salam,
``Change Of Variables And Equivalence Theorems In Quantum Field Theories,''
Nucl.\ Phys.\  {\bf 28} (1961) 529.

\bibitem{Bahns:2002vm}
D.~Bahns, S.~Doplicher, K.~Fredenhagen and G.~Piacitelli,
``On the unitarity problem in space/time noncommutative theories,''
Phys.\ Lett.\ B {\bf 533} (2002) 178
[arXiv:hep-th/0201222].

\bibitem{Rieffel} M.~Rieffel, ``Non-commutative tori---A case study of
  non-commutative differentiable manifolds,'' Contemp.\ Math.\ {\bf
    105} 191.

\bibitem{Krajewski:1998gq} T.~Krajewski, ``G\'eom\'etrie non
  commutative et interactions fondamentales,'' arXiv:math-ph/9903047.

\bibitem{ConnesRieffel} A.~Connes and M.~Rieffel, ``Yang-Mills for
  non-commutative two-tori,'' Contemp.\ Math.\ {\bf 62} (1987) 237.

\bibitem{Krajewski:1999ja}
T.~Krajewski and R.~Wulkenhaar,
``Perturbative quantum gauge fields on the noncommutative torus,''
Int.\ J.\ Mod.\ Phys.\ A {\bf 15} (2000) 1011
[arXiv:hep-th/9903187].

\bibitem{Gracia-Bondia:1987kw} J.~M.~Gracia-Bond\'{\i}a and
  J.~C.~V\'arilly, ``Algebras Of Distributions Suitable For Phase
  Space Quantum Mechanics. 1,'' J.\ Math.\ Phys.\ {\bf 29} (1988) 869.

\bibitem{Bars:2001dy} I.~Bars, ``Nonpertubative effects of extreme
  localization in noncommutative geometry,'' arXiv:hep-th/0109132.

\bibitem{Connes:2001ef} A.~Connes and M.~Dubois-Violette,
  ``Noncommutative finite-dimensional manifolds. I. Spherical
  manifolds and related examples,'' arXiv:math.qa/0107070.

\bibitem{Cho:1999sg} S.~Cho, R.~Hinterding, J.~Madore and
  H.~Steinacker, ``Finite field theory on noncommutative geometries,''
  Int.\ J.\ Mod.\ Phys.\ D {\bf 9} (2000) 161 [arXiv:hep-th/9903239].

\bibitem{Martinetti:2001fq} P.~Martinetti and R.~Wulkenhaar,
  ``Discrete Kaluza-Klein from scalar fluctuations in noncommutative
  geometry,'' J.\ Math.\ Phys.\ {\bf 43} (2002) 182
  [arXiv:hep-th/0104108].

\bibitem{Gonzalez-Arroyo:1983ac}
A.~Gonzalez-Arroyo and C.~P.~Korthals Altes,
``Reduced Model For Large N Continuum Field Theories,''
Phys.\ Lett.\ B {\bf 131} (1983) 396.

\bibitem{Langmann:2002ai} E.~Langmann, ``Interacting fermions on
  noncommutative spaces: Exactly solvable quantum field theories in
  2n+1 dimensions,'' arXiv:hep-th/0205287.

\bibitem{Wilson:1973jj}
K.~G.~Wilson and J.~B.~Kogut,
``The Renormalization Group And The Epsilon Expansion,''
Phys.\ Rept.\  {\bf 12} (1974) 75.

\bibitem{Polchinski:1983gv}
J.~Polchinski,
``Renormalization And Effective Lagrangians,''
Nucl.\ Phys.\ B {\bf 231} (1984) 269.

\bibitem{Griguolo:2001ez} L.~Griguolo and M.~Pietroni, ``Wilsonian
  renormalization group and the non-commutative IR/UV connection,''
  JHEP {\bf 0105} (2001) 032 [arXiv:hep-th/0104217].

\bibitem{Martin:1999aq} C.~P.~Mart\'{\i}n and D.~S\'anchez-Ruiz, ``The
  one-loop UV divergent structure of U(1) Yang-Mills theory on
  noncommutative $\mathbb{R}^4$,'' Phys.\ Rev.\ Lett.\ {\bf 83} (1999)
  476 [arXiv:hep-th/9903077].

\bibitem{Sheikh-Jabbari:1999iw} M.~M.~Sheikh-Jabbari,
  ``Renormalizability of the supersymmetric Yang-Mills theories on the
  noncommutative torus,'' JHEP {\bf 9906} (1999) 015
  [arXiv:hep-th/9903107].

\bibitem{Minwalla:1999px} S.~Minwalla, M.~Van Raamsdonk and
  N.~Seiberg, ``Noncommutative perturbative dynamics,'' JHEP {\bf
    0002} (2000) 020 [arXiv:hep-th/9912072].

\bibitem{Grosse:2000yy} H.~Grosse, T.~Krajewski and R.~Wulkenhaar,
  ``Renormalization of noncommutative Yang-Mills theories: A simple
  example,'' arXiv:hep-th/0001182.

\bibitem{Matusis:2000jf} A.~Matusis, L.~Susskind and N.~Toumbas, ``The
  IR/UV connection in the non-commutative gauge theories,'' JHEP {\bf
    0012} (2000) 002 [arXiv:hep-th/0002075].

\bibitem{Chepelev:1999tt} I.~Chepelev and R.~Roiban, ``Renormalization
  of quantum field theories on noncommutative $\mathbb{R}^d$.  I:
  Scalars,'' JHEP {\bf 0005} (2000) 037 [arXiv:hep-th/9911098].

\bibitem{Chepelev:2000hm} I.~Chepelev and R.~Roiban, ``Convergence
  theorem for non-commutative Feynman graphs and renormalization,''
  JHEP {\bf 0103} (2001) 001 [arXiv:hep-th/0008090].

\bibitem{Lowenstein:1975ps} J.~H.~Lowenstein, ``Convergence Theorems
  For Renormalized Feynman Integrals With Zero - Mass Propagators,''
  Commun.\ Math.\ Phys.\ {\bf 47} (1976) 53.

\bibitem{Epstein:gw} H.~Epstein and V.~Glaser, ``The Role Of Locality
  In Perturbation Theory,'' Annales Poincare Phys.\ Theor.\ A {\bf 19}
  (1973) 211.

\bibitem{Seiberg:1999vs} N.~Seiberg and E.~Witten, ``String theory and
  noncommutative geometry,'' JHEP {\bf 9909} (1999) 032
  [arXiv:hep-th/9908142].

\bibitem{Bichl:2001yf} A.~A.~Bichl, J.~M.~Grimstrup, H.~Grosse,
  E.~Kraus, L.~Popp, M.~Schweda and R.~Wulkenhaar, ``Noncommutative
  Lorentz symmetry and the origin of the Seiberg-Witten map,'' Eur.\ 
  Phys.\ J.\ C {\bf 24} (2002) 165 [arXiv:hep-th/0108045].

\bibitem{Bichl:2001cq} A.~Bichl, J.~Grimstrup, H.~Grosse, L.~Popp,
  M.~Schweda and R.~Wulkenhaar, ``Renormalization of the
  noncommutative photon self-energy to all orders via Seiberg-Witten
  map,'' JHEP {\bf 0106} (2001) 013 [arXiv:hep-th/0104097].

\bibitem{Bichl:2001nf}
A.~A.~Bichl, J.~M.~Grimstrup, L.~Popp, M.~Schweda and R.~Wulkenhaar,
``Perturbative analysis of the Seiberg-Witten map,''
Int.\ J.\ Mod.\ Phys.\ A {\bf 17} (2002) 2219
[arXiv:hep-th/0102044].

\bibitem{Wulkenhaar:2001sq} R.~Wulkenhaar, ``Non-renormalizability of
  $\theta$-expanded noncommutative QED,'' JHEP {\bf 0203} (2002) 024
  [arXiv:hep-th/0112248].

\bibitem{Grimstrup:2002af}
J.~M.~Grimstrup and R.~Wulkenhaar,
``Quantisation of $\theta$-expanded non-commutative QED,''
arXiv:hep-th/0205153.



\end{thebibliography}
\end{document}